# Physical Activity Trajectories Preceding Incident Major Depressive Disorder Diagnosis Using Consumer Wearable Devices in the All of Us Research Program: Case-Control Study

Yuezhou Zhang[1], PhD; Amos Folarin[1,2,3,4,5], PhD; Hugh Logan Ellis[1], MbBChir; Rongrong Zhong[1,6], MD; Callum Stewart[1], PhD; Heet Sankesara[1], BSc; Hyunju Kim[1], PhD; Shaoxiong Sun[1,7], PhD; Abhishek Pratap[1,8,9], PhD; Richard JB Dobson[1,2,3,4,5], PhD

[1]Department of Biostatistics & Health Informatics, Institute of Psychiatry, Psychology and Neuroscience, King's College London, London, United Kingdom

[2]Institute of Health Informatics, University College London, London, United Kingdom

[3]NIHR Biomedical Research Centre at South London and Maudsley NHS Foundation Trust, London, United Kingdom

[4]NIHR Biomedical Research Centre at University College London Hospitals NHS Foundation Trust, London, United Kingdom

[5]Health Data Research UK, University College London, London, United Kingdom

[6]Clinical Research Center & Division of Mood Disorders, Shanghai Mental Health Center, Shanghai Jiao Tong University School of Medicine, Shanghai, China

[7]Department of Computer Science, University of Sheffield, Sheffield, United Kingdom

[8]Boehringer Ingelheim Pharmaceuticals, Inc., Ridgefield, CT, United States

[9]University of Washington, Seattle, WA, United States

**Corresponding author:**

Richard JB Dobson, PhD

Department of Biostatistics & Health Informatics, Institute of Psychiatry, Psychology and Neuroscience, King's College London, United Kingdom

Mail address: PO Box 80, SGDP Centre, IoPPN, De Crespigny Park, Denmark Hill London, SE5 8AF, United Kingdom

Email: richard.j.dobson@kcl.ac.uk

Phone: 44 20 7848 0473

**Abstract**

**Background:** Low physical activity is a well-established risk factor for major depressive disorder (MDD). However, how physical activity changes in the months preceding a first clinical diagnosis of MDD has not been well explored, particularly when assessed using long-term, objective measures collected in real-world settings.

**Objective:** This study aimed to characterize trajectories of wearable-measured physical activity during the year preceding incident MDD diagnosis.

**Methods:** We conducted a retrospective nested case-control study using linked electronic health record (EHR) and consumer wearable data (Fitbit) from the All of Us Research Program. Adults with at least 6 months of valid Fitbit physical activity data in the 12 months preceding diagnosis were eligible. Incident MDD cases were identified based on a first recorded EHR diagnosis and matched to MDD-free controls on age, sex, body mass index, and time of diagnosis, with up to four controls per case. Daily step counts and moderate-to-vigorous physical activity (MVPA) were aggregated into monthly averages. Linear mixed-effects models were used to compare pre-diagnostic physical activity trajectories between cases and controls over a retrospective time scale from −12 to 0 months. Among cases, within-person contrasts were used to identify when physical activity levels first showed statistically significant deviations relative to levels observed 12 months before diagnosis. Exploratory analyses examined whether pre-diagnostic trajectories differed by sex, age, and body mass index.

**Results:** The analytic cohort included 4,104 participants (829 incident MDD cases and 3,275

matched controls; 81.7% women; median age 48.4 years). Compared with controls, individuals who developed MDD exhibited consistently lower overall physical activity levels and significant downward trajectories in both daily step counts and MVPA during the year preceding diagnosis (global trajectory tests, $P < .001$ for both outcomes). Among cases, statistically significant changes in daily step counts emerged approximately 4 months before diagnosis, whereas significant changes in MVPA emerged approximately 5 months before diagnosis. Furthermore, exploratory analyses suggested heterogeneity in pre-diagnostic trajectories across demographic subgroups, including steeper declines among men, more pronounced reductions in activity intensity at older ages, and persistently lower activity levels with flatter trajectories among individuals with obesity.

**Conclusions:** In this large cohort with linked wearable and clinical data, sustained within-person declines in physical activity emerged several months before incident MDD diagnosis. These findings suggest that longitudinal monitoring of physical activity using consumer wearable devices may provide clinically relevant early signals to support risk stratification, targeted prevention, and earlier intervention for MDD.

**Introduction**

Major depressive disorder (MDD) is a leading cause of disability worldwide [1] and is associated with substantial adverse outcomes, including premature mortality [2], functional impairment [3], increased medical comorbidity [4], and suicide [5]. Given its substantial individual and societal burden [1], early identification of individuals in the prodromal or initial stages of MDD is critically important, as preventive strategies [6] and early interventions [7] may be most effective during this period. However, symptom onset frequently precedes clinical recognition and formal diagnosis in routine clinical settings [8, 9], reflecting prolonged durations of untreated illness in MDD.

Low physical activity (PA) has been identified as a risk factor for MDD in prospective cohort studies and meta-analyses [10-12]. Longitudinal evidence further suggests that reductions in PA may precede worsening depressive symptoms [13], supporting the possibility that declines in PA can occur before the clinical diagnosis of MDD. However, much of the existing evidence relies on self-reported PA, which is subject to recall bias [11], or on accelerometer-based assessments conducted over short durations or at infrequent follow-up time points [14]. Consequently, current studies provide limited insight into the timing and pattern of PA changes preceding a clinical MDD diagnosis, including when PA deviates from an individual's prior activity level and whether these changes differ across population subgroups. Addressing these questions requires longitudinal, objective measures of PA collected in real-world settings and reliably linked to clinical diagnostic information, which allow behavioral change to be characterized prior to diagnosis as illness unfolds in routine care.

Advances in sensor technology and the widespread adoption of consumer wearable devices have enabled passive, continuous monitoring of real-world PA at scale, with devices such as Fitbit trackers capturing daily step counts and activity intensity—key correlates of functioning and psychomotor activity [15]—over extended periods with minimal user burden [16, 17]. Recent mobile health studies have demonstrated negative associations between wearable-measured PA and depression severity [18-20], suggesting that such measures may complement symptom-based assessments by informing risk stratification and prompting earlier clinical evaluation. Nevertheless, prior mobile health studies have been limited by modest sample sizes, relatively short monitoring durations, or the absence of linkage to clinical diagnoses [18], constraining their ability to characterize behavioral changes in relation to clinical diagnosis timing.

Few large-scale datasets integrate long-term wearable data with electronic health records (EHRs) in a manner that enables examination of pre-diagnostic PA trajectories and subgroup heterogeneity. The All of Us Research Program (AoURP) provides such an integrated resource, comprising a large, ongoing national cohort in which participants consent to share EHRs, health surveys, biospecimens, physical measurements, and wearable data [21]. By linking historical and prospective Fitbit data with EHR-based diagnoses, this study enables examination of long-term PA trajectories preceding clinical MDD diagnosis. Here, we aimed to determine whether PA trajectories differ between individuals with incident MDD and matched controls during the 12 months preceding diagnosis. Among individuals who developed MDD, we further examined when PA begins to deviate from earlier levels and whether the pattern of pre-diagnostic changes varies across population subgroups.

## Methods

### Study participants

We used data from the AoURP, an ongoing national longitudinal cohort funded by the US National Institutes of Health [21], with the long-term goal of enrolling at least 1 million participants. The study design and data collection procedures have been described previously [21, 22].

The present analysis used the controlled tier dataset, version 8 (C2024Q3R8), including participants enrolled between May 2017 and October 2023. Participant demographics and baseline data were collected during the digital enrollment. For participants who consented to share EHR and Fitbit data, their historical (pre-enrollment) EHR data and Fitbit data were made available through their participating health care provider organizations and linked Google Fitbit accounts, respectively [21-23]. In this data version, 36,614 individuals had linked EHR and Fitbit data available.

### Ethical Considerations

This study involved a secondary analysis of deidentified data obtained from the AoURP and therefore did not require additional ethics review. Access to deidentified data was restricted to authorized study investigators who completed required All of Us Responsible Conduct of Research training, and all analyses were conducted within the secure, cloud-based Researcher Workbench environment. In accordance with the AoURP Data and Statistics Dissemination Policy, analytic results for groups with fewer than 20 participants were not reported to minimize the risk of participant reidentification.

For the AoURP, all participants provided informed consent at enrollment and were informed of their right to withdraw from the program at any time. Data privacy and confidentiality were protected through multiple safeguards, including secure cloud-based data storage, restricted access to deidentified data, and mandatory confidentiality and data use agreements. Participant compensation of US $25 for the collection of biological specimens (eg, blood, saliva, or urine) was provided by the AoURP in the form of cash payments, gift cards, or electronic vouchers, as applicable.

**Fitbit PA Data**

PA in this study was quantified using daily step counts and moderate-to-vigorous physical activity (MVPA), which intuitively capture overall activity volume [24] and activity intensity emphasized in established public health guidelines [25], respectively. Fitbit classifies activity intensity into device-defined categories using metabolic equivalent of task (MET)–based criteria, consistent with intensity frameworks widely adopted in research-grade accelerometry [26, 27]. In this study, daily MVPA was calculated as the daily sum of device-labeled "fairly active" minutes plus twice the number of "very active" minutes, consistent with established definitions [8].

To ensure data quality, only days with at least 10 hours of wear time and daily step counts between 100 and 45,000 were considered valid, following prior AoURP studies [28-30]. Monthly averages of daily step counts and MVPA minutes were calculated to reduce day-to-day variability related to missingness and short-term fluctuations, consistent with prior AoURP analyses [28-30]. Months with more than 10 valid days were considered valid and retained for analysis [30].

**Incident MDD Cases**

Following definitions used in prior AoURP research [28, 31], incident MDD cases were defined as participants whose first recorded MDD diagnosis occurred during their Fitbit monitoring period, with no prior MDD diagnosis. Participants were required to be aged 19 years or older at the time of diagnosis to ensure they were adults (≥18 years) throughout the 1-year pre-diagnostic observation period. In addition, individuals with any recorded diagnosis of bipolar disorder, schizophrenia, or schizoaffective disorder prior to the diagnosis date were excluded. To ensure sufficient pre-diagnostic wearable data, cases were required to have at least 6 valid months of Fitbit PA data within the 12 months preceding the diagnosis month.

Diagnoses were identified from EHR data using standardized concept identifiers from the Observational Medical Outcomes Partnership (OMOP) common data model, which harmonize diagnoses across vocabularies including SNOMED, ICD-9-CM, and ICD-10-CM [32]. The OMOP concept identifiers used in this study, along with their corresponding ICD-9-CM and ICD-10-CM codes, are provided in Supplementary Table 1.

**Nested Case-Control Design**

A nested case-control design [33-35] was used to construct the comparison group. Cases and controls were matched at the diagnosis month (hereafter referred to as the matching month) on age (within 1 year), sex, and body mass index (BMI) category, which were selected a priori given their strong associations with both PA and depression risk. BMI was defined using the measurement closest to one year before the matching month to minimize potential influence of emerging MDD on body weight. Eligible controls were selected from the risk set of participants who, at the matching month, were aged 19 years or older, had no recorded MDD diagnosis, had no prior diagnosis of bipolar disorder, schizophrenia, or schizoaffective disorder, and met the same Fitbit PA data availability criteria as cases. For each case, up to four controls were selected, a ratio chosen to improve statistical efficiency while maintaining matching quality and consistency with prior trajectory-based nested case-control studies [33, 34]. When more than four eligible controls were available, controls were randomly sampled without replacement within or between cases to avoid repeated use of the same individuals and to simplify the correlation structure of longitudinal analyses [34]. Additional sociodemographic factors were not included in matching to avoid overmatching and preserve statistical efficiency.

**Statistical analysis**

Linear mixed-effects models with participant-specific random intercepts were used to estimate PA trajectories over time. A retrospective time scale from −12 to 0 months was defined, with time 0 corresponding to the diagnosis month for cases and the matching month for controls. Models included case–control status (coded as 0 for controls and 1 for cases), linear and quadratic time terms (time and time$^2$) to allow for potential non-linear temporal patterns, and their interaction terms, as well as matching variables. Differences in trajectories between cases and controls were evaluated using time×case and time$^2$×case interaction terms, with a joint Wald test used to evaluate whether overall temporal patterns differed by case-control status. Model-derived marginal means were used to estimate and compare monthly trajectories between cases and controls. Statistical significance of between-group differences at each month was assessed using contrast P values, with P values adjusted for multiple comparisons using the Benjamini–Hochberg procedure [36].

Among individuals with incident MDD, marginal means were estimated for each month, and pairwise contrasts comparing each month with month −12 were used to identify when PA began to differ relative to the earliest month. To assess potential heterogeneity in pre-diagnostic PA trajectories, exploratory case-only models were extended to include interactions between time variables and subgroup indicators for sex, age, and BMI.

**Results**

This study included 4,104 participants, comprising 829 individuals with incident MDD and 3,275 matched controls. The median age of the overall cohort was 48.4 years (IQR, 36.3–61.3 years); 81.7% of participants were women, and 82.5% were White. Cases and controls had similar distributions of age, sex, BMI, race, and ethnicity, with all standardized mean differences (SMDs) below 0.1, indicating adequate balance between groups (Table 1). Participant selection and cohort construction are shown in Figure 1.

**PA Trajectories Preceding Incident MDD Diagnosis**

The trajectories of daily step counts during the 12 months preceding the matching month differed significantly between incident MDD cases and matched controls (Figure 2A). In mixed-effects models, controls exhibited a relatively stable trajectory, whereas cases showed a marked and accelerating decline as the diagnosis month approached. These differences were supported by significant linear and quadratic interaction terms (time × case: $\beta = -92.08$; 95% CI, −126.14 to −58.02; time² × case: $\beta = -5.47$; 95% CI, −8.23 to −2.70), with a significant global test for trajectory differences (joint Wald test, $P < .001$) (Table 2). Model-derived marginal means indicated that daily step counts were consistently lower among cases than controls at every month examined. As early as 12 months before diagnosis, cases averaged 7568 steps (95% CI, 7330–7806) compared with 8360 steps (95% CI, 8240–8480) among controls (contrast difference, −792 steps; 95% CI, −1058 to −525; $P < .001$). This case–control gap widened progressively over time, reaching −1109 steps (95% CI, −1374 to −845; $P < .001$) in the matching month, when cases averaged 7140 steps (95% CI, 6904–7376) and controls averaged 8249 steps (95% CI, 8130–8369) (Supplementary Table 2).

Similar patterns were observed for daily MVPA trajectories (Figure 2B). Compared with controls, cases exhibited consistently lower daily MVPA levels throughout the pre-diagnostic period. In mixed-effects models, divergence in MVPA trajectories was supported by a significant linear interaction (time × case: $\beta = -0.78$; 95% CI, −1.39 to −0.18) and a significant global test for trajectory differences (joint Wald test, $P < .001$) (Table 2). Model-derived marginal means showed that at 12 months before diagnosis, cases averaged 52.9 minutes of MVPA (95% CI, 49.2–56.7) compared with 60.9 minutes (95% CI, 59.1–62.8) among controls (contrast $P < .001$). This difference increased over time and reached −11.35 minutes (95% CI, −15.48 to −7.22; $P < .001$) in the matching month, when cases averaged 47.3 minutes (95% CI, 43.7–51.0) and controls averaged 58.7 minutes (95% CI, 56.8–60.6) (Supplementary Table 3).

**Timing of Within-Person Pre-diagnostic Changes Among Incident MDD Cases**

To characterize the timing of within-person changes in PA preceding diagnosis, we conducted case-only marginal contrasts analyses using 12 months before diagnosis (time = −12) as the reference. Among cases, daily step counts did not differ significantly from the reference level until 4 months before diagnosis. At time = −4, step counts were significantly lower than at time = −12 (−145 steps; 95% CI, −253 to −37; *P* = .02). Reductions became larger as diagnosis approached, reaching −428 steps in the matching month (95% CI, −531 to −326; *P* < .001) (Supplementary Table 4).

In terms of daily MVPA, levels among cases were significantly lower than the reference level by 5 months before diagnosis (time = −5: −2.48 minutes; 95% CI, −4.32 to −0.64; *P* = .02). MVPA levels continued to decline over the pre-diagnostic period, reaching −5.61 minutes in the matching month (95% CI, −7.35 to −3.86; *P* < .001) (Supplementary Table 5).

**Subgroup Differences in Pre-Diagnostic PA Trajectories Among Cases**

Case-only mixed-effects models revealed significant heterogeneity in pre-diagnostic PA trajectories by sex, age, and BMI (Figure 3; Supplementary Tables 6–7). Men exhibited steeper declines in both daily step counts and MVPA than women, supported by significant linear and quadratic time-by-sex interaction terms (joint Wald tests, P < 0.001).

PA trajectories also varied significantly by age and BMI category (joint Wald tests, *P* < .001). Participants aged ≥60 years exhibited lower overall daily step levels and greater reductions in daily MVPA over time. Across BMI categories, overweight and obese participants had lower overall levels of daily steps and MVPA than those with normal BMI, and obese participants showed comparatively attenuated temporal declines in both outcomes (Figures 3E and 3F). The underweight group was excluded from subgroup analyses because of insufficient sample size.

**Discussion**

**Principal Findings**

In the AoURP cohort with linked wearable and EHR data, we found that adults who later received an incident MDD diagnosis exhibited a marked, progressively steepening decline in PA during the year preceding diagnosis compared with matched controls. Importantly, PA levels began to deviate from individuals' usual patterns approximately 4–5 months before clinical diagnosis, highlighting a clinically relevant window in which persistent within-person downward trajectories, rather than transient drops, may represent a more informative signal of emerging change. These findings suggest that longitudinal, objective PA monitoring may provide clinically interpretable insight into functioning and psychomotor behavior, although careful evaluation of potential confounding and subgroup heterogeneity is warranted.

Our findings extend prior evidence linking lower PA to subsequent MDD diagnosis. Prior prospective cohort studies [10-12, 14], including work in the AoURP cohort [28], have consistently associated lower PA levels with a higher risk of incident MDD, a pattern also observed in our analysis, with cases exhibiting lower PA than matched controls. Beyond these established differences, our analyses revealed two key patterns: cases showed significantly sustained declines in PA relative to controls, and, within cases, PA deviated from habitual levels approximately 4–5 months before diagnosis. Importantly, because EHR-documented diagnosis likely occurs after symptom onset and care-seeking [8, 9], the observed pre-diagnostic declines in PA should be interpreted with caution. These patterns may reflect prodromal processes [37-39], early symptomatic impairment, or a combination of both, rather than implying causal or temporal primacy of PA changes in the development of MDD.

Clinically, the primary implication of longitudinal, objective PA monitoring is not to diagnose MDD, but to provide an early, scalable pre-diagnostic signal of emerging vulnerability in routine care that may help trigger depression screening, preventive strategies, or early intervention. Our findings suggest a clinically relevant window in which sustained within-person declines in activity may be more informative than fixed activity thresholds, which are susceptible to transient fluctuations arising from a range of non-pathological factors. PA represents an actionable target across the course of depressive illness, as it is both a potentially modifiable risk factor for depression[6, 40, 41] and a well-established component of interventions that reduce depressive symptoms among individuals with established MDD[42, 43]. Accordingly, whether observed PA declines reflect prodromal vulnerability or early symptomatic impairment, timely prevention or intervention may confer clinical benefit[10, 11, 42-45], support more efficient use of health care resources, and reduce the broader societal burden of MDD [46, 47].

We also observed substantial heterogeneity in pre-diagnostic PA trajectories by sex, age, and BMI. Declines tended to be steeper among men, reductions in higher-intensity activity were more pronounced at older ages, and individuals with obesity exhibited persistently lower activity and flatter trajectories. Several factors may contribute to these patterns, including sex differences in baseline activity levels [48, 49], timing of help-seeking for mental health symptoms [50, 51], and social contexts of PA [52], as well as age- or weight-related functional limitations or comorbidities that constrain higher-intensity behaviors [53]. These interpretations remain exploratory and warrant further investigation.

**Limitations**

This study has several limitations. First, incident MDD was defined by the absence of EHR-documented diagnoses prior to Fitbit monitoring, consistent with prior AoURP methodologies [28, 30, 31], but this definition is primarily vulnerable to left-censoring of prior disease history and may not capture undiagnosed cases, warranting further confirmation in future studies. Second, observed pre-diagnostic declines in PA may be influenced by factors such as intercurrent illness, injury, cardiometabolic disease, or medication changes, which are not always fully captured or precisely timed in EHRs, limiting the ability to fully account for these influences in the present analyses and highlighting the need for future studies with explicit control of these covariates. Third, selection bias is also possible, as inclusion required participants to have sufficiently long Fitbit monitoring histories and meet prespecified data completeness criteria, potentially limiting representativeness. Fourth, the study population was predominantly White, and the bring-your-own-device nature of Fitbit data contribution may have enriched the analytic sample for more health-conscious or technologically engaged participants, potentially limiting generalizability.

**Conclusions**

In this large cohort with linked wearable and EHR data, adults who later received an incident diagnosis of MDD exhibited sustained, within-person declines in physical activity during the year preceding diagnosis, with deviations from habitual patterns emerging several months before clinical recognition and varying across demographic subgroups. Together, these findings suggest that longitudinal, objective PA monitoring using consumer wearables may offer clinically relevant early signals to support risk stratification, prevention, and earlier intervention for MDD, while underscoring the need for careful consideration of confounding and subgroup heterogeneity.

**Acknowledgments**

We gratefully acknowledge All of Us participants for their contributions, without whom this research would not have been possible. We also thank the National Institutes of Health's All of Us Research Program for making available the participant data examined in this study.

**Funding**

Richard JB Dobson is supported by the following: (1) National Institute for Health and Care Research (NIHR) Biomedical Research Centre (BRC) at South London and Maudsley National Health Service (NHS) Foundation Trust and King's College London; (2) Health Data Research UK, which is funded by the UK Medical Research Council (MRC), Engineering and Physical Sciences Research Council, Economic and Social Research Council, Department of Health and Social Care (England), Chief Scientist Office of the Scottish Government Health and Social Care Directorates, Health and Social Care Research and Development Division (Welsh Government), Public Health Agency (Northern Ireland), British Heart Foundation, and Wellcome Trust; (3) the BigData@Heart Consortium, funded by the Innovative Medicines Initiative 2 Joint Undertaking (which receives support from the EU's Horizon 2020 research and innovation programme and European Federation of Pharmaceutical Industries and Associations [EFPIA], partnering with 20 academic and industry partners and European Society of Cardiology); (4) the NIHR University College London Hospitals BRC; (5) the NIHR BRC at South London and Maudsley (related to attendance at the American Medical Informatics Association) NHS Foundation Trust and King's College London; (6) the UK Research and Innovation (UKRI) London Medical Imaging & Artificial Intelligence Centre for Value Based Healthcare (AI4VBH); (7) the NIHR Applied Research Collaboration (ARC) South London at King's College Hospital NHS Foundation Trust; and (8) Wellcome Trust. The funders had no role in the design and conduct of the study; collection, management, analysis, and interpretation of the data; preparation, review, or approval of the manuscript; and decision to submit the manuscript for publication.

**Conflicts of Interest**

Abhishek Pratap is an employee of Boehringer Ingelheim Pharmaceuticals, Inc., Ridgefield, CT, USA. Amos Folarin and Richard JB Dobson are cofounders of Onsentia. Amos Folarin holds shares of Google.

**Data Availability**

To ensure participant privacy, the data used in this study are available to approved researchers through the All of Us Research Workbench (https://workbench.researchallofus.org/login) following registration, completion of required ethics training, and attestation of a data use agreement. Public release of individual-level participant data is not permitted under the All of Us Research Program privacy policies.

**Authors' Contributions**

YZ had full access to all study data and takes responsibility for the integrity of the data and the accuracy of the data analysis. YZ, AF, and RJBD contributed to the study concept and design. Data acquisition, analysis, or interpretation were performed by YZ, AF, AP, HLE, RZ, CS, HS, HK, SS, and RJBD. YZ drafted the manuscript and conducted the statistical analysis.

Funding was obtained by RJBD and AF. All authors critically reviewed the manuscript for important intellectual content.

**Abbreviations**

AoURP: All of Us Research Program

BMI: body mass index

CI: confidence interval

EHR: electronic health record

ICD: International Classification of Diseases

IQR: interquartile range

MET: metabolic equivalent of task

MDD: major depressive disorder

MVPA: moderate-to-vigorous physical activity

OMOP: Observational Medical Outcomes Partnership

PA: physical activity

SMD: standardized mean difference

**Table 1. Characteristics of Incident Major Depressive Disorder Cases and Matched Controls.**

| Characteristics | Overall (N = 4104) | Case (n = 829) | Control (n = 3275) | SMD[a] |
|---|---|---|---|---|
| Age[b], median [IQR], y | 48.4 [36.3, 61.3] | 48.4 [36.2, 61.2] | 48.4 [36.3, 61.4] | 0.004 |
| BMI[c], median [IQR] | 29.9 [25.6, 34.9] | 29.8 [25.6, 35.4] | 29.9 [25.7, 34.7] | 0.046 |
| Women, n (%) | 3355 (81.7) | 676 (81.5) | 2679 (81.8) | 0.033 |
| Race, n (%) | | | | 0.098 |
|   White | 3384 (82.5) | 691 (83.4) | 2693 (82.2) | |
|   Black or African American | 271 (6.6) | 40 (4.8) | 231 (7.1) | |
|   Other[d] | 449 (10.9) | 98 (11.8) | 351 (10.7) | |
| Ethnicity, n (%) | | | | 0.025 |
|   Hispanic or Latino | 267 (6.5) | 56 (6.8) | 211 (6.4) | |
|   Not Hispanic or Latino[e] | 3837 (93.5) | 773 (93.2) | 3064 (93.6) | |

Abbreviations: BMI, body mass index; IQR, interquartile range; SMD, standardized mean difference.

[a]SMDs are presented as absolute values; values less than 0.1 indicate negligible imbalance between groups.

[b]Age was defined at the diagnosis month for cases and the matched index month for controls.

[c]BMI was derived from measurements closest to one year before the diagnosis or matching month to minimize potential influence of emerging major depressive disorder on body weight.

[d]Participants with missing, unknown, or less frequently reported race categories were grouped into the "Other" category in accordance with All of Us data reporting policies.

[e]Participants with missing or unknown ethnicity were grouped with the "Not Hispanic or Latino" category in accordance with All of Us data reporting policies.

**Table 2. Linear Mixed-Effects Model Comparing Pre-diagnostic Trajectories of Daily Step Counts and Moderate-to-Vigorous Physical Activity (MVPA) Between Incident MDD Cases and Controls.**

| Term | Estimate (95% CI) | Term $P$ value | Global Trajectory Test ($P$ Value)[a] |
|---|---|---|---|
| Daily step count model (unit: steps)[b,c,d,e] | | | <.001 |
|     Intercept | 8249.57 (8130.24 to 8368.90) | <.001 | |
|     group × case | −1109.57 (−1373.76 to −845.39) | <.001 | |
|     time | 3.66 (−11.58 to 18.90) | .64 | |
|     time × case | −92.08 (−126.14 to −58.02) | <.001 | |
|     time² | 1.07 (−0.16 to 2.30) | .09 | |
|     time² × case | −5.47 (−8.23 to −2.70) | <.001 | |
| MVPA model (unit: minutes) | | | <.001 |
|     Intercept | 58.69 (56.82 to 60.55) | <.001 | |
|     group × case | −11.35 (−15.48 to −7.22) | <.001 | |
|     time | 0.05 (−0.23 to 0.32) | .75 | |
|     time × case | −0.78 (−1.39 to −0.18) | .01 | |
|     time² | 0.02 (−0.00 to 0.04) | .08 | |
|     time² × case | −0.04 (−0.09 to 0.01) | .10 | |

[a] The global trajectory test (joint Wald test) evaluates whether overall physical activity trajectories differ between cases and controls by jointly testing the interaction terms (time × case and time² × case).
[b] Time was modeled on a retrospective monthly scale from −12 to 0 months relative to the diagnosis (or matching) month.
[c] The intercept represents the estimated mean physical activity level for the reference group (controls) at the reference time point (month 0).
[d] Group × case represents the baseline difference between incident major depressive disorder cases and matched controls (case–control status).
[e] Time × case and time² × case represent differential linear and quadratic changes trajectories between cases and controls over time.

**Figure 1. Study Flowchart for the Identification of Incident MDD Cases and Matched Controls**

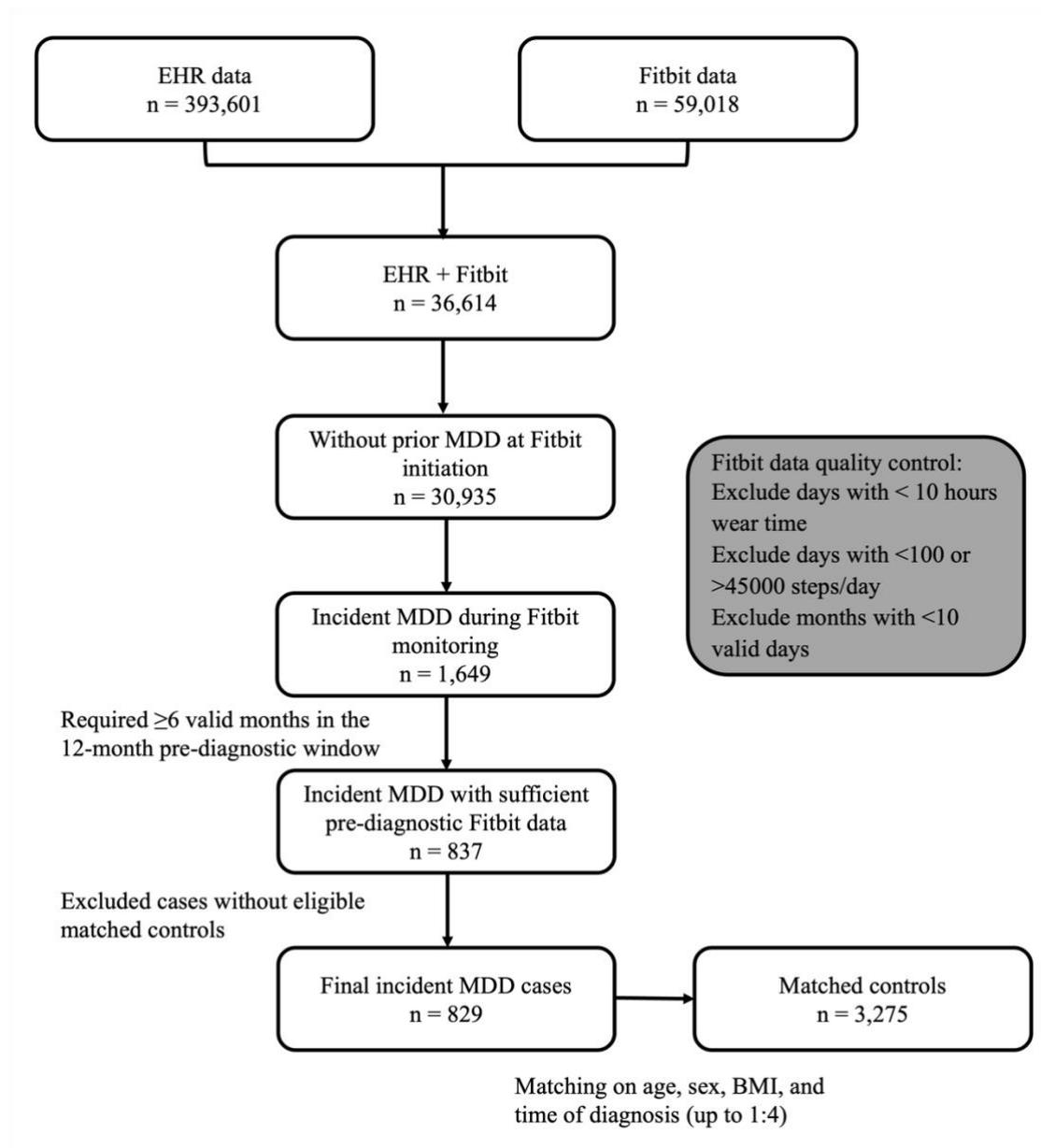

**Figure 2. Trajectories of Daily Step Counts (A) and Moderate-to-Vigorous Physical Activity (MVPA) (B) Preceding Incident MDD Diagnosis.**

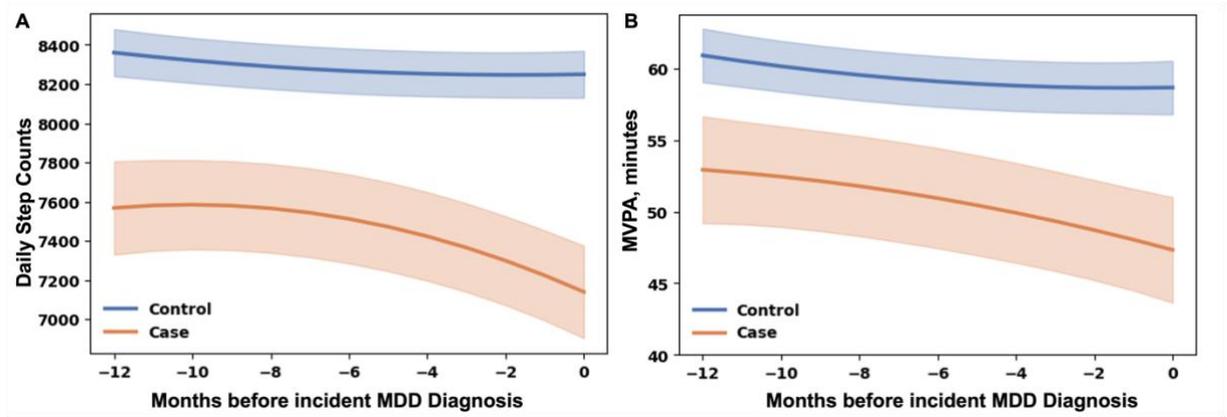

**Figure 3. Subgroup-Specific Trajectories of Physical Activity Preceding Incident MDD Diagnosis.** (A) Daily Step Count Trajectories by Sex. (B) Moderate-to-Vigorous Physical Activity (MVPA) Trajectories by Sex. (C) Daily Step Count Trajectories by Age. (D) MVPA Trajectories by Age. (E) Daily Step Count Trajectories by Body Mass Index (BMI). (F) MVPA Trajectories by BMI.

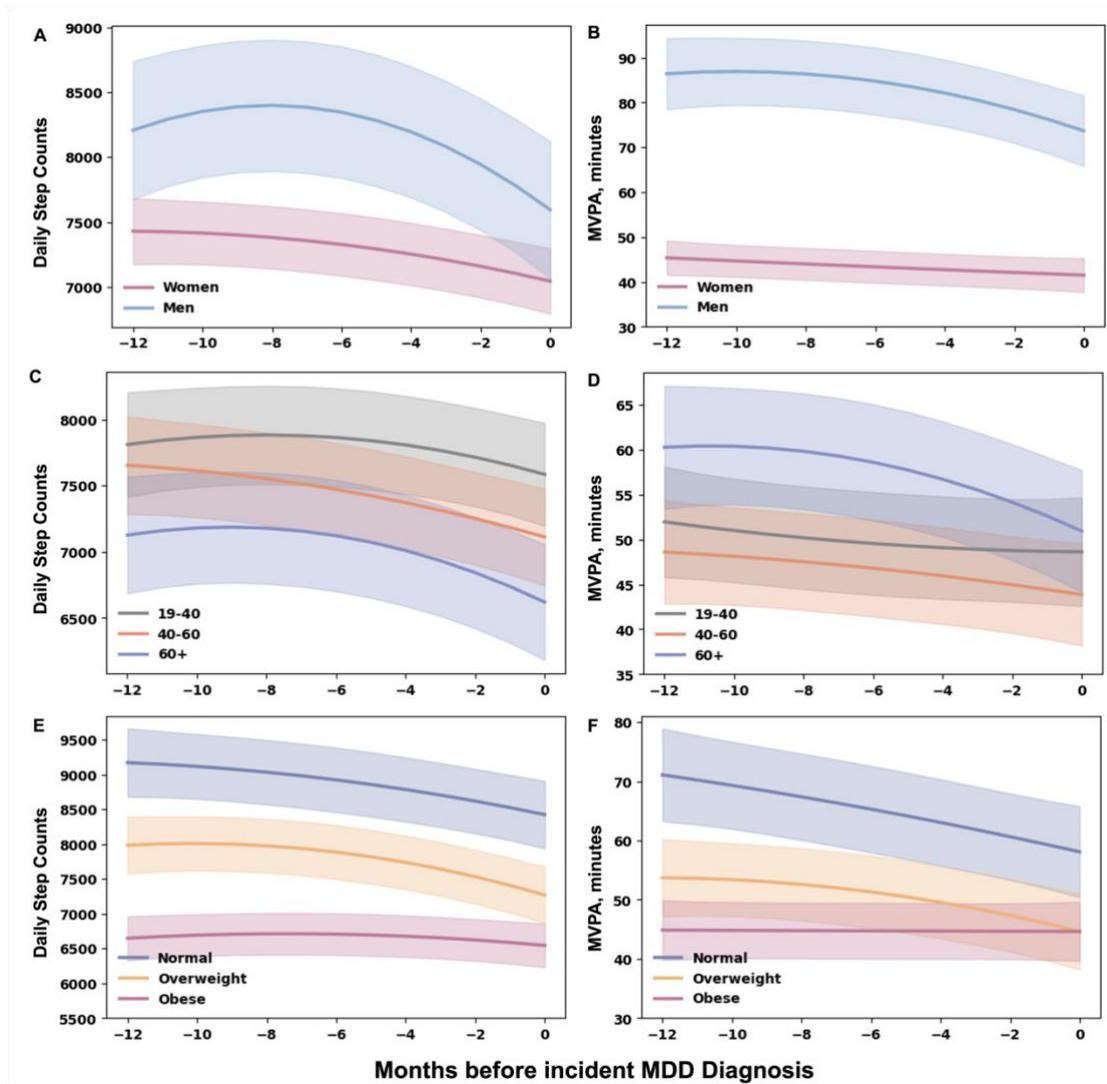